\documentclass[a4paper,12pt,authoryear]{elsarticle}
\usepackage{amsmath,amsfonts,amssymb,epsfig}
\usepackage{fullpage}
\usepackage{natbib}

\usepackage{color}

\usepackage{mathdots}
\usepackage{stmaryrd}

\newcommand{\be}{\begin{equation}}
\newcommand{\en}{\end{equation}}

\DeclareMathOperator{\tr}{tr}

\newcommand{\ii}{\textrm{i}}
\newcommand{\ee}{\textrm{e}}
\renewcommand{\vec}[1]{\boldsymbol{#1}}

\begin{document}

\begin{frontmatter}

\title{Counter-intuitive results in acousto-elasticity\footnote{Dedicated to V.I. Alshits}}
\author[label1]{A.L. Gower}
\author[label1,label2]{M. Destrade}
\author[label3]{R.W. Ogden}

\address[label1]{School of Mathematics, Statistics and Applied Mathematics,\\
National University of Ireland Galway,\\
University Road, Galway, Ireland}
\address[label2]{School of Mechanical {\&} Materials Engineering, \\
University College Dublin, \\
Belfield, Dublin 4, Ireland}
\address[label3]{School of Mathematics and Statistics, \\
University of Glasgow, Scotland}

\date{\today}

\begin{abstract}

We present examples of body wave and surface wave propagation in deformed solids where the slowest and the fastest waves do not travel along the directions of least and greatest stretch, respectively. These results run counter to commonly accepted theory, practice, and implementation of the principles of acousto-elasticity in initially isotropic solids.
For instance we find that in nickel and steel, the fastest waves are along the direction of greatest compression, not greatest extension (and vice-versa for the slowest waves), as soon as those solids are deformed.
Further, we find that when some materials are subject to a small-but-finite deformations, other extrema of wave speeds appear in non-principal directions. Examples include nickel, steel, polystyrene, and a certain hydrogel.
The existence of these ``oblique'', non-principal extremal waves complicates the protocols for the non-destructive determination of the directions of extreme strains.

\end{abstract}


\end{frontmatter}

\numberwithin{equation}{section}


\section{Introduction}

\begin{figure}[!h]
\centerline{\includegraphics[height=0.3\textheight]{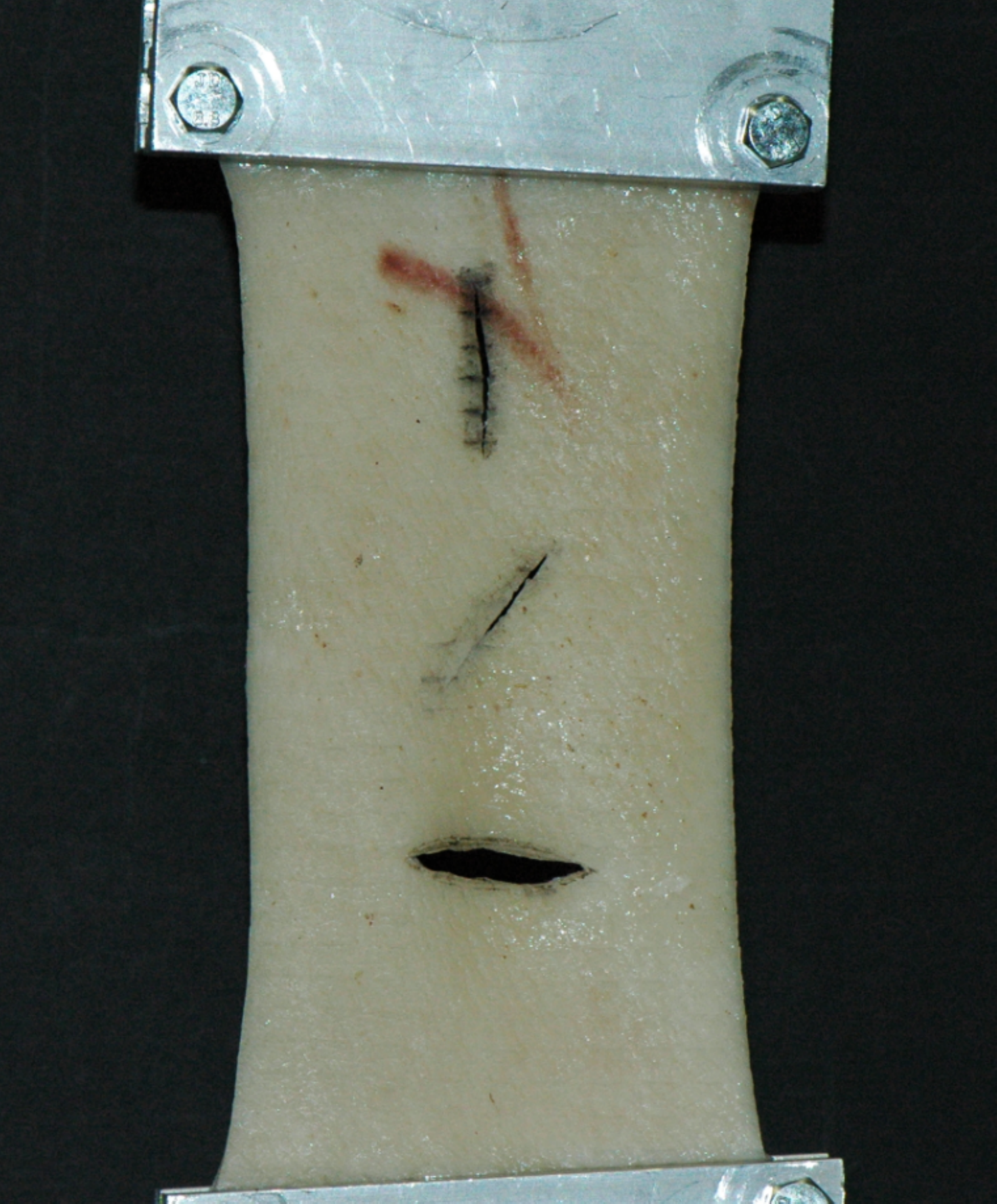}} \caption{Cutting through pig skin: a clamped sample of pig skin is put under tension, after 3 cuts have been performed, parallel (top), oblique (center) and perpendicular (bottom) to the tensile force.
} \label{fig1}
\end{figure}

The determination of the direction of greatest tension in a deformed solid is one of the main goals of acoustic non-destructive evaluation because, for isotropic solids, this direction coincides with the direction of greatest stress.
Consider for instance cutting through a membrane under uniaxial tension: cutting parallel to the direction of the tensile force produces a thin cut, while cutting across produces a gaping cut (see Figure \ref{fig1}), which can have serious consequences in scaring outcomes after stabbing incidents or surgery.
Finding the direction of greatest stress is also important in geophysics, oil prospecting \citep{GuJo09} and structural health monitoring and evaluation \citep{PaSa84, KiSa01}.

In this paper we investigate the propagation of small-amplitude elastic waves in the body (body acoustic waves -- BAWs) and on the surface (surface acoustic waves -- SAWs) of a deformed solid, and determine the dependence of their speeds on the angle of propagation with respect to the principal directions of pre-strain.
It is widely thought that surface waves propagate at their fastest in the direction of greatest stretch and at their slowest in a perpendicular direction, along the direction of least stretch.  This view is supported by intuition and often forms the basis of a non-destructive determination of these directions.

However, the coupling of acoustics and elasticity is a \emph{non-linear phenomenon} even at its lowest order, and it can thus generate counter-intuitive results.
The first such result is that for some materials, the fastest wave travels along the direction of greatest compression (and conversely, the slowest wave along the direction of greatest extension).  It has been known for some time that a compression in one direction could indeed result  in an increase in the speed of a principal wave instead of the intuitively expected decrease,  and \cite{HuKe53} showed experimentally that body wave speeds increase with hydrostatic pressure for polystyrene (see their Figure 3); similar experimental results exist for body waves in railroad steel and surface waves in mild steel; see \cite{KiSa01} for a review.
Here we extend those results to the consideration of non-principal waves in deformed steel and nickel, and to pre-strains resulting in turn from the application of a uniaxial stress and of a pure shear stress.

The other counter-intuitive result is that the following statement by \cite{KiSa01} is not necessarily true: ``The principal stress direction is found where the variations of the SAW speeds show symmetry about the direction''.
Indeed, \cite{Tanu12} recently showed that for a small-amplitude SAW traveling in the symmetry plane of a transversely isotropic solid, subject to a small pre-strain, the correction to the wave speed due to the pre-stress has sinusoidal variations with respect to the angle of propagation, in line with that statement.
Explicitly, \cite{Tanu12} established the following expression for the correction to the Rayleigh wave speed $v^0_R$ when the solid is subject to a pre-stress with principal components $\sigma_1$, $\sigma_2$ in the plane boundary:
\[
v_R = v_R^0 + A (\sigma_1+\sigma_2) + B(\sigma_1-\sigma_2)\cos2\psi,
\]
where $A$ and $B$ are acousto-elastic coefficients, and $\psi$  is the angle between the direction of propagation and one of the principal directions of pre-stress.
However, their result is only true when the pre-stress and accompanying pre-strain are infinitesimal. Here we show that the variations can  rapidly lose their sinusoidal regularity beyond that regime, even when a solid is deformed by as little as 1\%.
Since in non-destructive evaluation and structural health monitoring, the order of magnitude of the pre-stress is not known \emph{a priori}, we conclude that a complete theoretical and numerical investigation needs to be conducted (as here) prior to the determination of the sought-after principal directions.
They will not be found simply by measuring the wave speed in all directions until a symmetry in variation is found.

The paper is organized as follows.
In the next section we recall the equations governing the propagation of small-amplitude waves in solids subject to a pre-strain of arbitrary magnitude.
For the constitutive modeling, we focus on isotropic solids with a strain-energy density expressed as a polynomial expansion up to third order in terms of invariants of the Green strain tensor.
Historically, this is the framework in which the equations of \emph{acousto-elasticity} have often been written  in considering elastic wave propagation in a slightly pre-deformed, initially isotropic solid.
We refer to, for example, \cite{PaSa84} and \cite{KiSa01} for an exposition of the practical and theoretical aspects of this technique, which can be dated back to the early efforts of  \cite{Bril25} and \cite{HuKe53}; see also \cite{DeOg13} for a review of acousto-elasticity in solids subject to a general homogeneous pre-strain (not necessarily of infinitesimal amplitude).
In Sections \ref{Body waves section} and \ref{Surface waves section}, we study body wave and surface wave  propagation, respectively (the latter is more complicated than the former, and we thus devote Section \ref{Riccati section} to a description of our numerical strategy).
For both types of waves we uncover examples of solids (steel, Pyrex glass, polystyrene, nickel, hydrogel with a hard core) where the wave speed does not have its greatest value along the direction of greatest stretch, and/or can be extremal along directions which are \emph{oblique} to the directions of the principal stretches.
These counter-intuitive results seem to have gone unnoticed before.


\section{Governing equations}
\label{Section2}


\subsection{Incremental motions}
\label{Section2.1}

In this paper we are concerned with the propagation of small-amplitude waves in deformed materials.  The equations governing their motion are now well established.
Consider a homogeneous elastic solid, held in a state of static homogeneous deformation, which has brought a material point which was at $\vec{X}$ in the reference configuration to position $\vec{x} = \vec{x}(\vec{X},t)$ in the current configuration.

Let $(X'_1,X'_2,X'_3)$ be the coordinates of $\vec{X}$ with respect to a fixed rectangular Cartesian unit basis vectors $(\vec{e}_1', \vec{e}_2',\vec{e}_3')$, and let a pure homogeneous strain be defined by
\begin{equation}
x_1'=\lambda_1 X'_1,\quad x_2'=\lambda_2 X'_2,\quad x_3'=\lambda_3 X'_3,
\end{equation}
with respect to the same basis, where the positive constants $\lambda_1,\lambda_2,\lambda_3$ are the \emph{principal stretches} of the deformation.  Now consider the material to be a half-space occupying the region $x'_2\geq 0$ so that the boundary $x'_2=0$ is a principal plane of deformation, which we take to be free of traction.  Now choose a second set of unit basis vectors $(\vec{e}_1, \vec{e}_2,\vec{e}_3)$, say, with coordinates ($x_1,x_2,x_3$), so that
$x_2=x_2'$ and the direction of $\vec{e}_1$ makes an angle $\theta$ with the direction of $\vec{e}_1'$. Then
\begin{equation} \label{eqns:Deformation}
\begin{bmatrix}
x_1 \\ x_2 \\ x_3 \end{bmatrix}
 = \begin{bmatrix} \cos \theta & 0 & \sin \theta \\ 0 & 1 & 0 \\ -\sin \theta & 0 & \cos \theta
  \end{bmatrix}
  \begin{bmatrix}
x_1' \\ x_2' \\ x_3' \end{bmatrix}.
\end{equation}

A small-amplitude wave traveling in this material is described by the associated mechanical displacement field $\vec{u} = \vec{u}(\vec{x},t)$, satisfying, in the coordinate system $(x_1,x_2,x_3)$, the incremental equations of motion \citep{Ogde97},
\begin{equation}
\label{eqn:IncrementalMotion}
\rho u_{i,tt} = s_{pi,p}= \mathcal A_{0piqj} u_{j,pq},
\end{equation}
where $s_{pi}=\mathcal A_{0piqj} u_{j,q}$ are the components of the incremental nominal stress tensor, and $\mathcal{A}_{0piqj}$ are components of  the fourth-order tensor of instantaneous moduli $\vec{\mathcal{A}}_0$ (to be detailed later), a comma followed by an index $i$ (or $t$) denotes partial differentiation with respect to $x_i,\,i=1,2,3$, (or $t$) and $\rho$ is the current mass density.
We specialize the analysis to waves that propagate in the $x_1$ direction, with amplitude variations in the $x_2$ direction. Hence we seek solutions of the form
\begin{equation} \label{eqn:WaveForm}
\vec{u} = \vec{U}(x_2) \ee^{\ii k (x_1-vt)},
\end{equation}
where $\vec{U}$, the amplitude, is a function of $x_2$ only, $k$ is the wavenumber, and $v$ is the wave speed.
Then the equations of motion reduce to
\begin{equation} \label{ode:U}
\vec{\sf{T}}\vec{U}''(y) + \ii k(\vec{\sf{R}}+\vec{\sf{R}}^\mathrm{T})\vec{U}'(y) - k^2(\vec{\sf{Q}}-\rho v^2 \vec{\sf{I}})\vec{U}(y) = \vec{0},
\end{equation}
where the constant tensors $\vec{\sf{T}}$, $\vec{\sf{R}}$, $\vec{\sf{Q}}$ are defined in terms of their components with respect to the basis $(\vec{e}_1, \vec{e}_2,\vec{e}_3)$ by
\begin{equation} \label{TQR}
 T_{ij} = T_{ji}  = \mathcal{A}_{02i2i}, \quad R_{ij} = \mathcal{A}_{02i1j}, \quad Q_{ij} = Q_{ji} = \mathcal{A}_{01i1j},
\end{equation}
$\vec{\sf{I}}$ is the identity tensor, and the exponent $^\mathrm{T}$ denotes the transpose.

Without loss of generality, we take $\lambda_1 <  \lambda_2< \lambda_3$, so that $\theta = 0^\circ$ corresponds to the direction of  greatest compression and $\theta = 90^\circ$ to the direction of greatest stretch.
In the coordinate system ($x'_1,x'_2,x'_3$) aligned with the principal axes of deformation, there are only 15 non-zero components of $\vec{\mathcal{A}_0}$, given by \citep{Ogde97}
\begin{align}
&  \mathcal{A}'_{0iijj} = J^{-1}\lambda_i \lambda_j W_{ij}, & \notag \\
&  \mathcal{A}'_{0ijij} = J^{-1}(\lambda_i W_i - \lambda_j W_j)\lambda_i^2/(\lambda_i^2-\lambda_j^2), & i\neq j, \; \lambda_i \neq \lambda_j,  \notag \\
& \mathcal{A}'_{0ijji} = J^{-1} (\lambda_j W_i - \lambda_i W_j)\lambda_i \lambda_j/(\lambda_i^2-\lambda_j^2),& i\neq j, \; \lambda_i \neq \lambda_j,  \notag \\
&  \mathcal{A}'_{0ijij} = J^{-1}(\lambda_i^2W_{ii} -\lambda_i\lambda_jW_{ij} + \lambda_iW_i)/2, & i\neq j, \; \lambda_i = \lambda_j,  \notag \\
& \mathcal{A}'_{0ijji} = \mathcal A'_{ijji} =  J^{-1} (\lambda_i^2 W_{ii} - \lambda_i\lambda_j W_{ij} - \lambda_i W_i)/2,& i\neq j, \; \lambda_i = \lambda_j,
\label{eqns:A}
\end{align}
where $J=\lambda_1\lambda_2\lambda_3$ is the dilatation, $W$ is the strain energy density, $W_i = \partial W/\partial \lambda_i$, $W_{ij} = \partial^2W/\partial \lambda_i \partial\lambda_j$ and there is no sum on repeated indices.
In the coordinate system ($x_1,x_2,x_3$), the components of $\vec{\mathcal{A}_0}$, required to compute the tensors in \eqref{TQR}, are given by
\begin{equation} \label{eqn:A0}
\mathcal{A}_{0ijkl} = \Omega_{ip}\Omega_{jq}\Omega_{kr}\Omega_{ls}\mathcal{A}'_{0pqrs},
\end{equation}
where $\Omega_{ij}$ is the rotation matrix corresponding to a rotation through the angle $\theta$ about $x_2=x_2'$.

We say that $ \vec {\mathcal{A}_0}$ satisfies
the \emph{strong-convexity} condition (S-C) when
\begin{equation} \label{cond:StrongConvexity}
\mathcal{A}_{0ijkl} \xi_{ij}\xi_{kl} > 0 \quad \mbox{for all non-zero tensors  } \vec{\xi},
\end{equation}
but we remark that this condition does not hold in general, only in the region of deformation space corresponding to dead-load stability (see, for example, \citealp{Ogde97}).
The \emph{strong-ellipticity} condition (S-E) reads
\begin{equation} \label{cond:StrongEllipticity}
\mathcal{A}_{0ijkl} n_i n_k m_j m_l > 0 \quad \mbox{for all non-zero vectors } \vec{n} \mbox{ and } \vec{m},
\end{equation}
and is implied by strong convexity.

\subsection{Deformed materials}
\label{Section2.2}

For the constitutive modeling of the pre-deformed materials, we focus on  general isotropic compressible elastic solids, with  a third-order expansion of the strain-energy density  in powers of the Green strain tensor $\vec{\sf{E}}$,  specifically
\begin{equation} \label{eqns:ThirdOrder}
W = \dfrac{\lambda_0}{2}i_1^2 + \mu_0i_2 +\dfrac{A}{3}i_3 + Bi_1i_2 +\dfrac{C}{3}i_1^3,
\end{equation}
where
\begin{equation}
i_k = \tr\left(\vec{\sf{E}}^k\right) = \dfrac{1}{2^k}\left[\left(\lambda_1^2-1\right)^k + \left(\lambda_2^2-1\right)^k + \left(\lambda_3^2-1\right)^k \right], \qquad k=1,2,3.
\end{equation}
Here, $\lambda_0$ and $\mu_0$ are the Lam\'e coefficients of second-order elasticity
and $A$, $B$, $C$ are the Landau coefficients of third-order elasticity \citep{LaLi86}.

For our examples, we use material parameters taken from the literature  for nickel \citep{Lurie}, steel \citep{Lurie}, polystyrene \citep{HuKe53}, Pyrex glass \citep{Lurie}, and a certain hydrogel with a hard core \citep{WuKi10} all summarized in Table 1.

\begin{table}[ht!]
\centering
\begin{tabular}{l | l | c c c c c }
Material  & Units & $\lambda_0$ & $\mu_0$ & $A$ & $B$ & $C$ \\
\hline
Nickel & $10^{5}$bars & 7.8 & 6.12857 & $ -73$ & $ -22.5$ & $17.9$  \\
Steel & $10^{5}$bars & 8.1 & 5.4 & $ -76$ & $ -25$ & $-9$  \\
Polystyrene & $10^5$bars & 0.2889 & 1.381 & $-1.00$ & $-0.830$ & $-1.06$  \\
Pyrex Glass & $10^5$bars & 2.75 & 5.583 & $42$ & $71$ & $-69.6$  \\
Hydrogel& NkT & 4595 & 1184 & $-2737$ & $-1682.5$ & $-3762.5$  \\
\hline
\end{tabular}
\caption{Second- and third-order elastic constants for six different materials.}
\end{table}

We look at two types of pre-deformations: first, that due to a uniaxial stress and second that due to a pure shear stress.
A \emph{uniaxial pre-stress} in the $\vec{e}_1'$ direction is due to a Cauchy stress for which the only non-zero component is $\sigma_{11} = T$, say.
It leads to an {equibiaxial pre-deformation}, with corresponding principal stretches
\begin{equation}
\label{eqn:UniaxialStrech}
\lambda_1 =  \lambda, \quad \lambda_2 =   \lambda_3.
\end{equation}
Here $\lambda$ is linked to the compressive stress $T$ through the equation $T = J^{-1} \lambda_1\partial W/\partial \lambda_1$, whilst $\lambda_2$ is found in terms of $\lambda$ by solving
\begin{equation}
\label{eqn:UniaxialStress}
0= \partial W/\partial \lambda_2.
\end{equation}
With our choice \eqref{eqns:ThirdOrder} of strain energy density, this turns out to be a quadratic in $\lambda_2^2$.

A \emph{pure shear stress} is applied parallel to the plane of the boundary so that the only non-zero Cauchy stress component is $\sigma_{13} = S$, say. The corresponding pre-deformation is a combination of  simple shear in the $x_3$ direction and a triaxial stretch \citep{MiGo11,DeMS12}.
Here it is a simple exercise to check (see, for example,  \citealp{Lurie}) that the principal stresses are $S$, $0$,  $-S$, and that the corresponding principal directions of stress are along $(1, 0, 1)$, $(0, 1, 0)$, $(1, 0, -1)$, respectively.
The principal directions of strain are aligned with these directions, and the principal stretches are found by solving the system
\begin{equation}
\label{eqn:ShearStress}
S = J^{-1}\lambda_1\partial W/\partial \lambda_1, \quad
0  = \partial W/\partial \lambda_2, \quad
-S = J^{-1} \lambda_3\partial W/\partial \lambda_3,
\end{equation}
for $\lambda_1 =\lambda$, $\lambda_2$ and $\lambda_3$.

The range of realistic values for $\lambda$ is restricted by the existence of a solution of the system of equations \eqref{eqn:UniaxialStrech} and \eqref{eqn:UniaxialStress} for uniaxial compression, and of the equations in \eqref{eqn:ShearStress} for pure shear stress.
There is a great variability of this feasible range for $\lambda$  from one material to another.
For example, steel can only be sheared for $\lambda$ from 1 down to $0.935$, below which value there are no real solutions, while for the  hydrogel there exists a pure shear stress solution for deformations of at least  40\%.
We further restrict our range of admissible $\lambda$'s by assuming that the materials are subject to uniaxial compressive stresses or pure shear stresses only within the region where $S$ and $T$ are monotone functions of $\lambda$. This ensures that our results belong to a physically valid regime.


\section{Results for body waves}
\label{Body waves section}


For homogeneous \emph{body} waves, there are no boundary conditions to satisfy and no amplitude variation to consider.
Hence we take
\begin{equation}
\label{def:PlaneBodyWave}
\vec{U}(x_2) = \vec{U}_0,
\end{equation}
a constant vector, in the governing equation (\ref{ode:U}), resulting in the eigenvalue problem
\begin{equation}
\label{eqn:SecularPlaneBodyWave}
 (\vec{\sf{Q}} -\rho v^2 \vec{\sf{I}}) \vec{U}_0 = \vec{0},
\end{equation}
with associated characteristic equation $\det(\vec{\sf{Q}}-\rho v^2\vec{\sf{I}})=0$, a cubic in $\rho v^2$.

For the body waves traveling along the principal direction corresponding to the least principal stretch $\lambda_1$, i.e. $\theta =0^\circ$, we find the three roots
\begin{equation}
\label{eqns:BodyWaveVelocity1}
\rho v^2 = \mathcal{A}'_{01111}, \ \mathcal{A}'_{01212},\ \mathcal{A}'_{01313},
\end{equation}
and similarly for the body waves along the principal direction corresponding to the largest stretch ratio $\lambda_3$, i.e. $\theta =90^\circ$,
\begin{equation}
\label{eqns:BodyWaveVelocity2}
\rho v^2 = \mathcal{A}'_{03333}, \ \mathcal{A}'_{03131}, \ \mathcal{A}'_{03232}.
\end{equation}
In each set of three roots for $\rho v^2$, the first root corresponds to a pure longitudinal wave and the next two to pure transverse waves.

In general ($\theta \ne 0, 90^\circ$), the characteristic equation factorizes into the product of a  term linear in $\rho v^2$ (corresponding to a pure transverse wave polarized along the $x_2$ direction) and a  term quadratic in $\rho v^2$ (with one root corresponding to a pseudo-longitudinal  wave and the other to a pseudo-transverse wave); see \cite{Norr83} for details.

\begin{figure}[!t]
\centerline{
a)
\includegraphics[height=0.22\textheight]{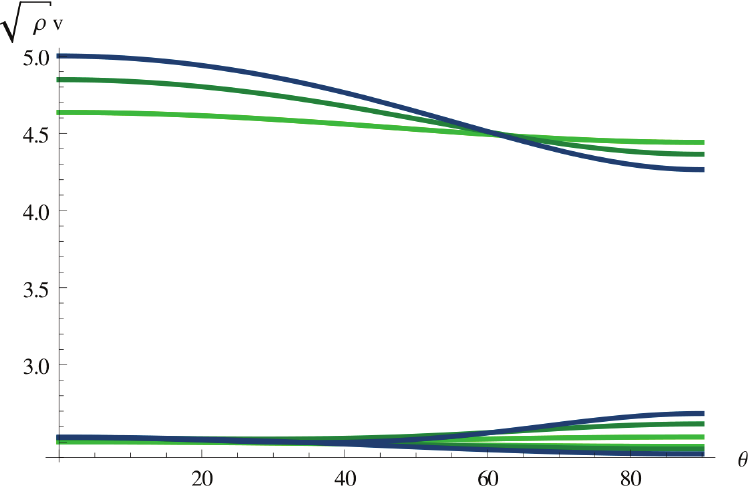}
b)
\includegraphics[height=0.22\textheight]{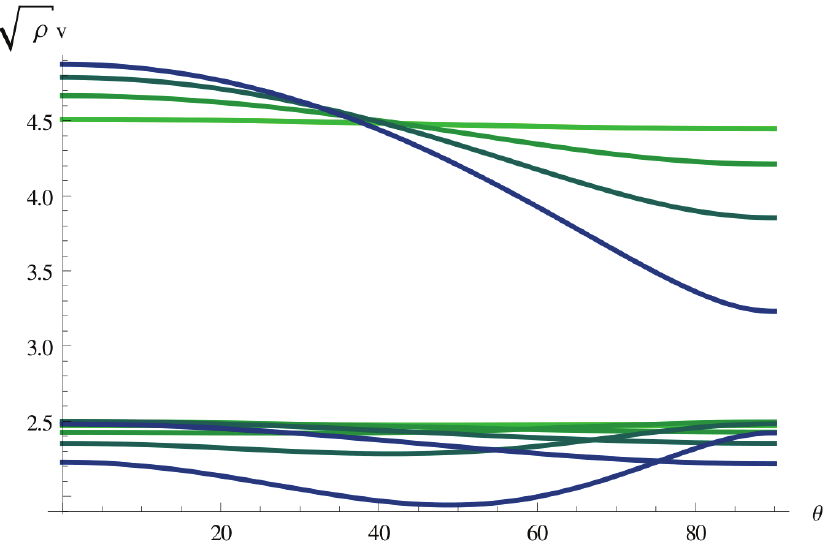}
}
\caption{The three body-wave speed profiles (plotted as $v\sqrt{\rho}$) for nickel under (a) uniaxial compressive stress with principal compression stretch ratio $\lambda =0.99$ (light green curve), $0.973$ (blue-green curve), $0.956$ (blue curve); (b) pure shear stress with $\lambda =0.99$ (light green curve), $0.978$ (blue-green curve) and $0.967$ (blue curve).}
\label{fig:nickelBodyWaves}
\end{figure}

Figure \ref{fig:nickelBodyWaves} depicts the variations of the three body wave speeds (in this and all subsequent plots it is $v\sqrt{\rho}$ that is plotted) in deformed \emph{nickel} with respect to the angle $\theta$ between the direction of greatest compression and the direction of propagation, for different values of compressive stretch under uniaxial and pure shear stresses.
The variations of the wave traveling with the intermediate speed meet intuitive expectations: this wave travels at its slowest when $\theta = 0^\circ$ and at its fastest when $\theta = 90^\circ$.
However, this scenario is reversed for the fastest and slowest waves, as soon as the solid is deformed: they travel at their fastest along the direction of greatest compression ($\theta = 0^\circ$) and slowest in the orthogonal direction.
Moreover, when a pure shear stress induces a compression of more than 3\%, we notice that the profile for the slowest body wave develops a new minimum; in effect this wave travels at its slowest in a direction which is \emph{oblique} with respect to the principal directions of strain ($\theta \simeq 50^\circ$).

\begin{figure}[!ht]
\centerline{
a)
\includegraphics[height=0.22\textheight]{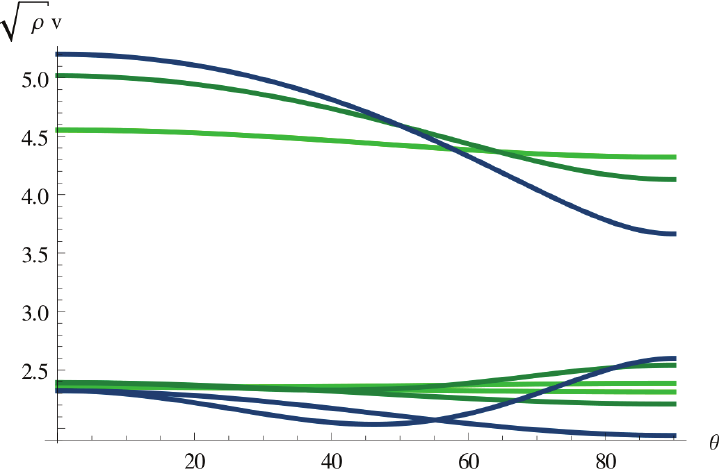}
b)
\includegraphics[height=0.22\textheight]{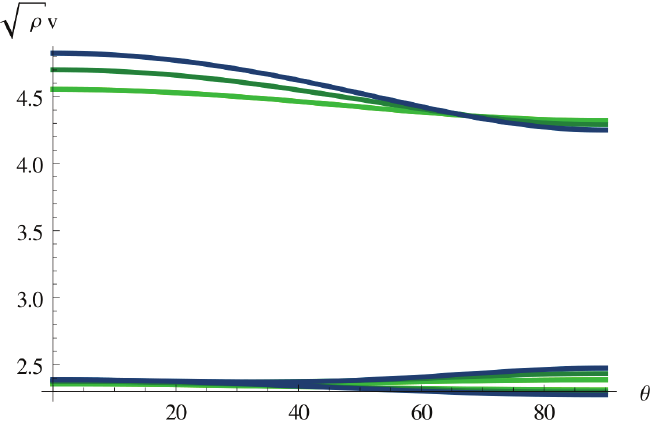}
}
\caption{The three body-wave speed profiles (plotted as $v\sqrt{\rho}$) for steel under (a)  uniaxial  compressive stress with $\lambda =0.99$ (light green curve), $0.956$ (blue-green curve), $0.922$ (blue curve); (b) pure shear stress with $\lambda =0.99$ (light green curve), $0.981$ (blue-green curve), $0.973$ (blue curve).}
\label{fig:steelBodyWaves}
\end{figure}

Figure \ref{fig:steelBodyWaves} show the corresponding results for deformed \emph{steel}.
They are similar to those for deformed nickel, with the difference that the secondary minimum phenomena occurs under uniaxial compression instead of pure shear stress.

\begin{figure}[!t]
\centerline{
a)
\includegraphics[height=0.22\textheight]{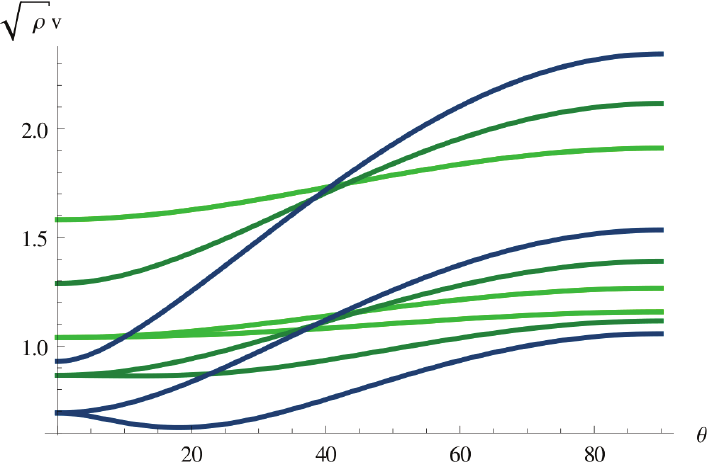}
b)
\includegraphics[height=0.22\textheight]{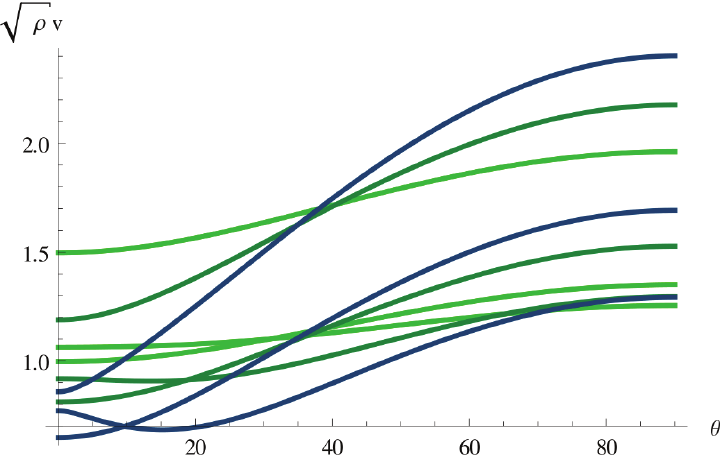}
}
\caption{The three body-wave speed profiles (plotted as $v\sqrt{\rho}$) for polystyrene under (a) uniaxial compressive stress; (b) pure shear stress. The light green curves correspond to $\lambda =0.91$, the blue-green curves to $\lambda=0.8$, and the blue curves to $\lambda=0.69$.}
\label{fig:PolyBodyWaves}
\end{figure}

\begin{figure}[!ht]
\centerline{
a)
\includegraphics[height=0.22\textheight]{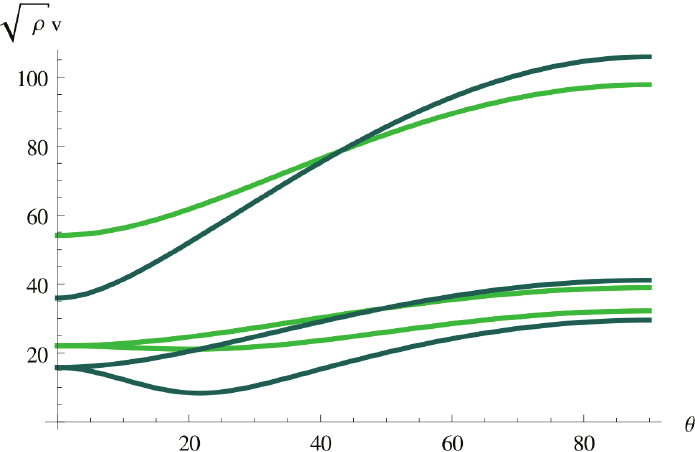}
b)
\includegraphics[height=0.22\textheight]{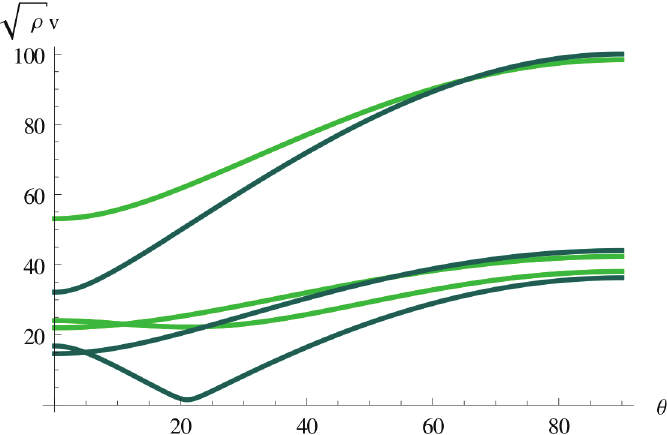}
}
\caption{The three body-wave speed profiles (plotted as $v\sqrt{\rho}$) for hydrogel under (a) uniaxial compressive stress; (b) pure shear stress. The light green curves correspond to $\lambda =0.75$ (i.e. 25\% maximum compression) and the dark green curves to $\lambda=0.61$ (i.e. 39\% maximum compression).}
\label{fig:WuBodyWaves}
\end{figure}

In Figures \ref{fig:PolyBodyWaves} and \ref{fig:WuBodyWaves}, we study body wave propagation in deformed \emph{polystyrene} and \emph{hydrogel}.
Here the waves all travel at their fastest along the direction of greatest stretch ($\theta = 90^\circ$) and two of the three waves travel at their slowest in the direction of greatest compression ($\theta = 0^\circ$).
There is, however, one wave which travels at its slowest in an oblique direction, for both types of pre-deformations (due to uniaxial stress: figures on the left; due to pure shear stress: figures on the right).
They appear at quite large compressions (31\% for polystyrene, 39\% for hydrogel), which are nonetheless compatible with the soft nature of these solids and with a physically acceptable material response (i.e. the tension and the shear stress are monotone functions of the stretch).

Now we investigate non-principal \emph{surface} wave propagation
in a deformed homogeneous half-space.
There are several methods of resolution available for these problems; see, for example, \cite{RoSa99, DOPR05, KWK11, GaML12}.
Here we adopt a formulation in terms of the surface impedance matrix.
In the next section we detail the steps involved in implementing this method, based on the analysis of \cite{FuMi02}.


\section{The matrix Riccati method for surface waves}
\label{Riccati section}


In the following, we replace the tensors $\vec{\sf{T}}$, $\vec{\sf{R}}$, $\vec{\sf{Q}}$, etc., introduced in the previous section by their matrix representations with respect to the Cartesian coordinates $(x_1,x_2,x_3)$.
In a nutshell, surface wave propagation is governed by the \emph{algebraic matrix Riccati equation} \citep{Biry85, BaLo85, FuMi02, NoSh10}
\begin{equation} \label{eqn:Riccati}
\mathsf{0} = [\mathsf{Z}(v)-\ii \mathsf{R}^\mathrm{T}]\mathsf{T}^{-1}[\mathsf{Z}(v)+\ii \mathsf{R}] - \mathsf{Q} + \rho v^2 \mathsf{I},
\end{equation}
the radiation condition,
\begin{equation} \label{cond:Decay}
\text{Im Spec } \mathsf{T}^{-1}[ \ii \mathsf{Z}(v) -\mathsf{R}] > 0,
\end{equation}
and the boundary condition of zero incremental traction on $x_2=0$, which is equivalent to
\begin{equation} \label{detZ}
\det \mathsf{Z}(v) =0.
\end{equation}
Here, the constant $3\times 3$ matrix $\mathsf{Z}(v)$ is the so-called \emph{surface impedance matrix}. For a given $v$, $\mathsf{Z}(v)$ is a constant Hermitian matrix, of the form
\begin{equation}
\mathsf{Z} = \begin{bmatrix} Z_1 & Z_4 + \ii Z_5 & Z_6 - \ii Z_7 \\
Z_4 - \ii Z_5 & Z_2 & Z_8+\ii Z_9 \\
Z_6 + \ii Z_7 & Z_8 - \ii Z_9 & Z_3 \end{bmatrix},
\end{equation}
say, where the $Z_k$ are real constants ($k=1, \ldots, 9$). The algebraic matrix Riccati equation \eqref{eqn:Riccati} is itself Hermitian, and thus corresponds to 9 real equations. Coupling it to \eqref{detZ} gives a system of 10 equations for the 10 unknowns $Z_k$ and $v$, and uniqueness of the solution comes from further requiring that $\mathsf{Z}(v)$ be \emph{positive definite}, as discussed below.

The surface impedance matrix $\mathsf{Z}(v)$ in a half-space relates the incremental displacement $\vec{u}$ to the incremental traction $\vec{t}$ on the surface $x_2 =$ constant through the relationship,
\begin{equation}
\label{eqn:ImpedanceRelation}
\vec{t} = -k \mathsf{Z}(v) \vec{u}.
\end{equation}
We may rewrite this by noting that the general solution of the homogeneous system of second-order ordinary differential equations with constant coefficients \eqref{ode:U} for the half-space, is of the form $\vec{U} = \ee^{\ii k \mathsf{E}(v) x_2} \vec{U}_0$, where $\mathsf{E}(v)$ is a constant $3 \times 3$ matrix (not to be confused with the Green strain) and $\vec{U}_0$ is a constant vector.
Then the traction is given by
\begin{equation} \label{imp-relation}
t_{i}= s_{2i}=\mathcal{A}_{02iqj}u_{j,q} ,
\quad \mbox{or} \quad
\vec{t}  = \ii k[\mathsf{R} + \mathsf{T} \mathsf{E}(v)] \vec{u}.
\end{equation}
Now write $\vec{t} =  - \ii k \vec{V} \ee^{\ii (k x_1 - v t)}$, where $\vec{V} = -[\mathsf{R} + \mathsf{T} \mathsf{E}(v)]\vec{U}$, so that the impedance relation \eqref{imp-relation} reads
\begin{equation}
\label{eqn:ImpedanceRelationReduced}
\vec{V}= - \ii\mathsf{Z}(v) \vec{U}, \quad \mbox{with} \quad \mathsf{Z}(v) = - \ii [\mathsf{R} + \mathsf{T} \mathsf{E}(v)],
\end{equation}
showing that $\mathsf{Z}(v)$ is indeed a constant matrix for a half-space.
The matrix $\mathsf{Z}(v)$ corresponding to the existence of a surface wave is the one that satisfies the Riccati equation \eqref{eqn:Riccati}, the boundary condition \eqref{detZ}, and 
\begin{equation*}
\text{Im Spec } \mathsf{E}(v) >0,
\end{equation*}
or, equivalenty, \eqref{cond:Decay}. This condition guarantees the correct decay for $\vec{U}(x_2) = \ee^{\ii k \mathsf{E}(v) x_2} \vec{U}_0$ as $x_2$ increases with distance away from the free surface.

In the matrix Riccati method, at least two remarkable properties emerge: $\mathsf{Z}(0)$ is positive definite in the region of stability and $\partial \mathsf{Z}(v)/\partial v$ is negative definite as long as \text{Im Spec }$\mathsf{T}^{-1}[ \ii \mathsf{Z}(v) -\mathsf{R}] > 0$.
Hence, $\det\mathsf{Z}(v)$ is positive at $v =0$ and monotonically decreasing as $v$ increases, which means that it is  simple to find $\tilde{v}$ numerically such that $\det \mathsf{Z}(\tilde{v}) =0$. Moreover uniqueness of the surface velocity, calculated by this procedure, is guaranteed.
\cite{BaLo85}, \cite{FuMi02} and \cite{MiFu03} have shown these properties, and here we present a somewhat simpler alternative demonstration
(see also \cite{ShPD04, AlMa05} for further impedance formulations).

Recall that the incremental nominal stress has components $s_{pi} = \mathcal{A}_{0piqj}u_{j,q}$ (with respect to the non-principal axes) and that the balance of momentum \eqref{eqn:IncrementalMotion} reads
\begin{equation}
\rho u_{i,tt} =  s_{pi,p}.
\end{equation}
Now multiply both sides of this by $u_i^*$, the complex conjugate of $u_i$:
\begin{equation}
\rho u^*_i u_{i,tt} = u^*_i s_{ji,j} = \left(u^*_i s_{ji} \right)_{,j} -  u^*_{i,j} s_{ji}\quad \mbox{with summation over $i$ and $j$}.
\end{equation}
Then integrate over the region $\mathcal{U} = [x_1,x_1+\Delta x_1] \times [0,\infty] \times [x_3,x_3+\Delta x_3]$ in the body, to obtain
\begin{equation}
\int_{\mathcal{U}} \rho u^*_i u_{i,tt}\, \mathrm{d}x_1\, \mathrm{d}x_2\, \mathrm{d}x_3 = \int_{\partial \mathcal{U}} u^*_i s_{ji} n_j\, \mathrm{d}a - \int_{ \mathcal{U}}  u^*_{i,j} s_{ji}\, \mathrm{d}x_1\, \mathrm{d}x_2\, \mathrm{d}x_3,
\end{equation}
where $\vec{n}$ is the outward unit normal vector to the boundary $\partial \mathcal{U}$ and $\mathrm{d}a$ the associated area element.
Now substitute $\vec{u}(x_1,x_2,x_3) = \vec{U}(x_2) \ee^{\ii k(x_1-vt)}$ to arrive at\footnote{Here and in the following we write the scalar product of two vectors as $\vec{a}\cdot\vec{b}$ rather than in the matrix form $\vec{a}^\mathrm{T}\vec{b}$.}
\begin{equation}
\label{eqn:StressPower0}
- k^2 v^2 \int_{0}^\infty  \rho \, \vec{U}^*(y)  \cdot \vec{U}(y)\, \mathrm{d}y =  u_i^* t_i \Big|_{y = 0}^{y = \infty}   - \int_{0}^\infty  \mathcal{A}_{0jilk} u_{i,j}^* u_{l,k}\, \mathrm{d}y ,
\end{equation}
 where we have introduced the components $t_i $, defined in \eqref{imp-relation}, of the traction $\vec{t}$ on planes normal to the $x_2$-axis.
 Observe that the above equation is independent of $x_1$ and $x_3$. Finally, assume that the wave amplitude decays away from the free surface, so that $\vec{U}(\infty) = \vec{0}$.
 Then substitute for $t_i$ from \eqref{eqn:ImpedanceRelation}  and rearrange to obtain
\begin{equation}
\label{eqn:StressPower}
 k\, \vec{U}^*(0) \cdot  \mathsf{Z}(v) \vec{U}(0) = \int_0^\infty  \mathcal{A}_{0jilk} u_{i,j}^* u_{l,k} \, \mathrm{d}x_2 - k^2 v^2  \int_{0}^\infty  \rho \,\vec{U}^*(x_2) \cdot \vec{U}(x_2)\, \mathrm{d}x_2.
\end{equation}

Here, only $\mathsf{Z}$ depends on $v$  because (i) $\vec{U}$ can be chosen independently of $v$ since for any choice of displacement field $\vec{U}$, a traction field $\vec{V}$ can be determined by equation~(\ref{eqn:ImpedanceRelationReduced}) such that momentum is balanced, and (ii)  $v$ cancels out in the products $u_{i,j}^* u_{l,k}$.
Therefore, by differentiating  with respect to $v$, we obtain
\begin{equation}
\label{eqn:dvZNegativeDefinite}
 \vec{U}^*(0)\cdot  \frac{\mathrm{d}\mathsf{Z}(v)}{\mathrm{d} v} \vec{U}(0) = - 2 k v  \int_{0}^\infty  \rho\, \vec{U}^*(x_2)\cdot \vec{U}(x_2)\, \mathrm{d}x_2 <0,
\end{equation}
while writing equation~(\ref{eqn:StressPower}) at $v=0$ gives
\begin{equation}
\label{eqn:Z0PositiveDefinite}
k\, \vec{U}^*(0) \cdot  \mathsf{Z}(0) \vec{U}(0) = \int_{0}^\infty \mathcal{A}_{0jilk} u_{i,j}^* u_{l,k} \,\mathrm{d}x_2,
\end{equation}
for any choice of $\vec{U}(0)$. Clearly  $\mathrm{d}\mathsf{Z}/\mathrm{d}v$ is negative definite by \eqref{eqn:dvZNegativeDefinite} and, from the strong-convexity condition (\ref{cond:StrongConvexity}) and \eqref{eqn:Z0PositiveDefinite}, $\mathsf{Z}(0)$ is positive definite if at least one of the components of $ u_{i,j}$ is non-zero.
Below we  show that $\mathsf{Z}(0)$ is positive definite when the deformation is within the region of (dead-load) stability.

For a \emph{material in the reference configuration}, strong-convexity is considered to be a necessary physical requirement, and it  implies that $\mathsf{Z}(0)$  is positive definite and that the decay condition \eqref{cond:Decay} holds at $v=0$. For a \emph{pre-stressed material}, strong-convexity is not expected in general. However, $\mathsf{Z}(0)$ is positive definite for a deformation in the region of dead-load stability.
Let the magnitude of the finite deformation be parameterized by $\alpha$, with $\alpha =0$ corresponding to no deformation (for instance, $\alpha$ can be the amount of shear in a simple shear pre-deformation, or the elongation $\lambda-1$ in a uniaxial stretch). Then the surface-impedance $\mathsf{Z}$ depends on $\alpha$ as well as on $v$ and the boundary condition of no incremental surface-traction (the secular equation) takes the form
\begin{equation}
\label{eqn:BucklingCriteria}
\det \mathsf{Z}(v,\alpha) =0.
\end{equation}
Assume that for $\alpha =0$ the strong-convexity condition~(\ref{cond:StrongConvexity}) is satisfied, so that $\mathsf{Z}(0,0)$ is positive definite. As $\alpha$ is increased and the deformation moves into the region of dead-load stability consider the change in the eigenvalues of $\mathsf{Z}(0,\alpha)$; these eigenvalues are positive until $\alpha$ reaches a critical value $\alpha^*$, say, when at least one eigenvalue becomes zero and $\det \mathsf{Z}(0,\alpha^*) =0$. At this point the half-space supports a standing-wave solution given by~(\ref{eqn:WaveForm}) with $v =0$ (at the boundary of the dead-load stability region), and the material has buckled (that is, it is unstable, at least in the linearized sense). For waves along the principal direction, this buckling criterion can be shown to be the same as found in \cite{DoOg91}.
For $\alpha > \alpha^*$ we say that the half-space is unstable with respect to surface-wave perturbations \citep{FuMi02}.

We are only interested in surface waves in the stable region $0 < \alpha < \alpha^*$ where $\mathsf{Z}(0,\alpha)$ is positive definite, and we define an implicit curve $v \to \mathsf{Z}(v,\alpha)$ by using the Riccati equation~(\ref{eqn:Riccati}). As long as Im Spec $\mathsf{T}^{-1}( \ii \mathsf{Z}(v,\alpha) -\mathsf{R}) > 0$ holds, we increase $v$ until $\det \mathsf{Z}(v,\alpha)=0$. If along this curve Im Spec $\mathsf{T}^{-1}( \ii \mathsf{Z}(v,\alpha) -\mathsf{R}) \leq 0$ before $\det \mathsf{Z}(v,\alpha)$ reaches zero, then there is no surface-wave.


\section{Results for surface waves}
 \label{Surface waves section}


We  transform the above analysis into a numerical method by choosing $\vec{\mathcal{A}}_0$ for which there is a positive definite $\mathsf{Z}(0)$ satisfying equation~(\ref{eqn:Riccati}).
Then, as $v$ is increased, we calculate the implicit curve for $\mathsf{Z}(v)$ from $\mathsf{Z}(0)$ up to $\mathsf{Z}(\hat{v})$ where $\det \mathsf{Z}(\hat{v}) =0$, all the while verifying that Im Spec $\mathsf{T}^{-1}( \ii \mathsf{Z} -\mathsf{R}) > 0$.
From that point on, we calculate another implicit curve that satisfies equations~(\ref{eqn:Riccati}) and \eqref{detZ} by varying $\vec{\mathcal{A}}_0$ (for instance, by varying the angle of propagation with respect to the principal axes or by varying the amplitude of the pre-deformation). If at some point $\mathsf{T}^{-1}( \ii \mathsf{Z} -\mathsf{R}) \leq 0$, then to confirm that there is no surface-wave calculate the implicit curve for $v \mapsto \mathsf{Z}(v)$ that departs from $\mathsf{}Z(0)$ and if, for some $v$, $\mathsf{T}^{-1}( \ii \mathsf{Z}(v) -\mathsf{R}) \leq 0$, then no surface-wave exists; if not, then varying $\vec{\mathcal{A}}_0$ has caused a discontinuous jump in the velocity, which may indeed be possible.

Using this method we now present Surface Acoustic Wave (SAW) velocity profiles in several materials subject to  either a uniaxial compressive stress or a pure shear stress, applied in the plane parallel to the free surface $x_2=0$.

\begin{figure}[!ht]
\centerline{
\includegraphics[height=0.22\textheight]{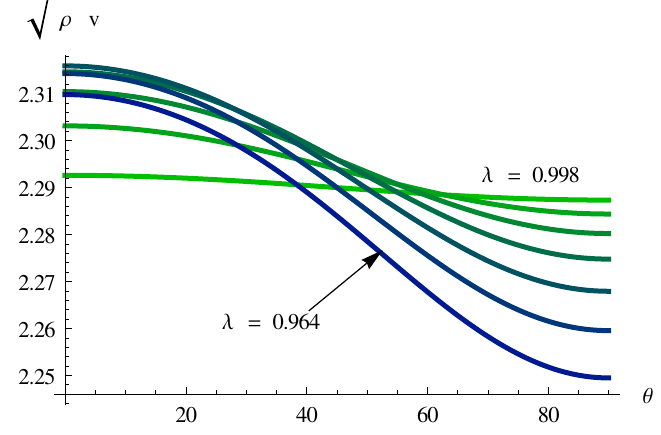}
\includegraphics[height=0.22\textheight]{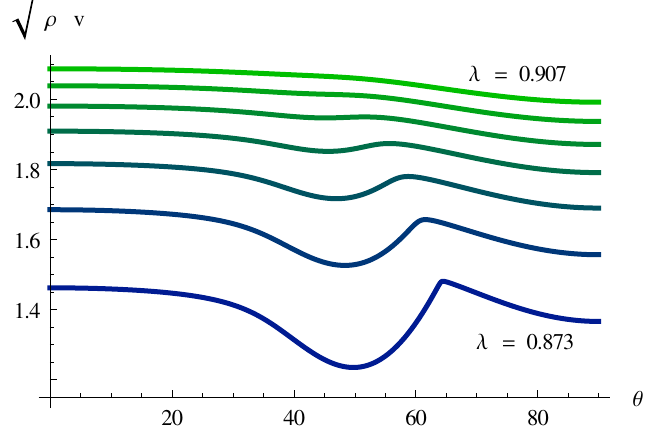}
}
\caption{Speed profiles for surface waves (plotted as $v\sqrt{\rho}$) in nickel subject  to uniaxial compressive stress, with pre-stretch $\lambda$ decreasing from 0.998 to 0.964 (on the left) and from 0.907 to 0.873 (on the right). As the color of the curves changes from green to blue, $\lambda$ is decreased by regular increments of 0.0057 from one curve to the next.}
\label{fig:nickelUniaxial}
\end{figure}

\begin{figure}[!ht]
\centerline{
\includegraphics[height=0.22\textheight]{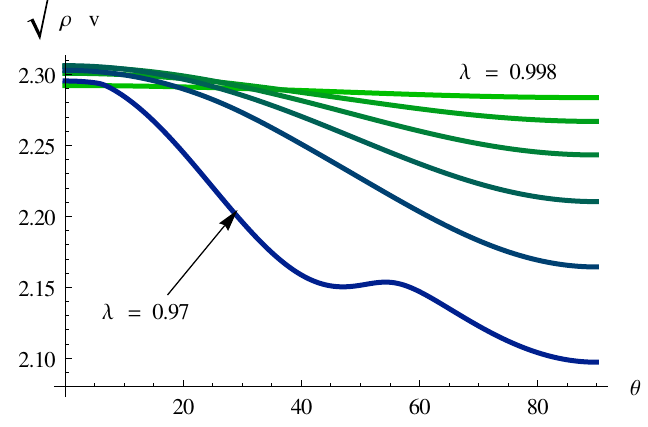}
\includegraphics[height=0.22\textheight]{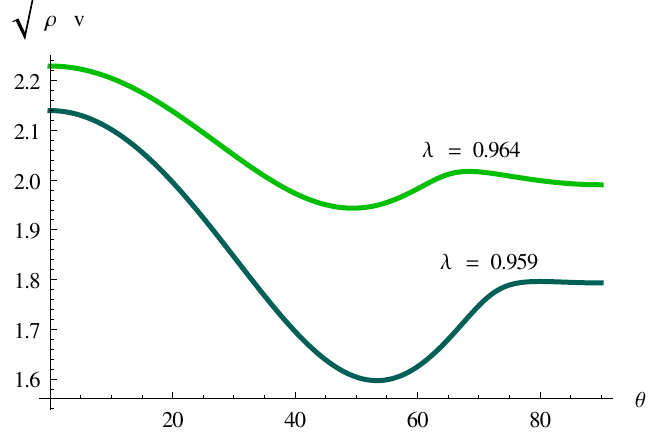}
}
\caption{Speed profiles for surface waves (plotted as $v\sqrt{\rho}$) in nickel subject  to pure shear stress, with pre-stretch $\lambda$ decreasing from 0.998 to 0.970 (on the left) and from 0.964 to 0.959 (on the right).
As the color of the curves  changes from green to blue, $\lambda$ is decreased by regular increments of 0.0057 from one curve to the next.}
\label{fig:nickelPureShear}
\end{figure}

Figure~\ref{fig:nickelUniaxial} depicts the variations of the surface wave speed with the angle of propagation with respect to the principal directions of strain in \emph{nickel} subject to a uniaxial compressive stress.
In the early stages of compression, from 1\% to 3\% compressive stretch say, ``the variations of the SAW speeds show symmetry about the [principal] direction[s]'' as stated by \cite{KiSa01}, with the proviso that the SAW travels at its fastest  along the direction of greatest compression $ \theta= 0^\circ$ and at its slowest along the direction of greatest stretch $ \theta= 90^\circ$ (in line with the behavior of the body waves in nickel, as shown in the previous section).
However, as the material is further compressed (compression beyond 10\%), secondary extrema develop: for $\lambda \geq 0.895$, the fastest SAW travels in the $\theta \simeq 65^\circ$ direction and the slowest SAW travels in the $\theta \simeq 45^\circ$ direction.
A similar phenomenon occurs when nickel is subject to a pure shear stress, as shown in Figure \ref{fig:nickelPureShear}: then the slowest wave travels at the oblique angle $\theta \simeq 50^\circ$ when the material is compressed by as little as 3.6\%; see figure on the right.

\begin{figure}[!ht]
\centerline{
a)
\includegraphics[height=0.22\textheight]{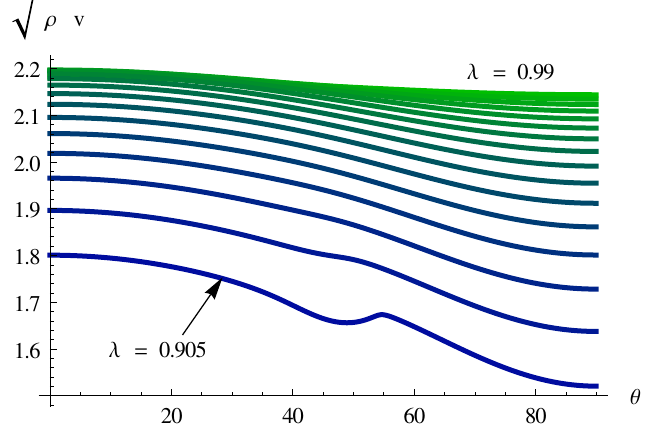}
b)
\includegraphics[height=0.22\textheight]{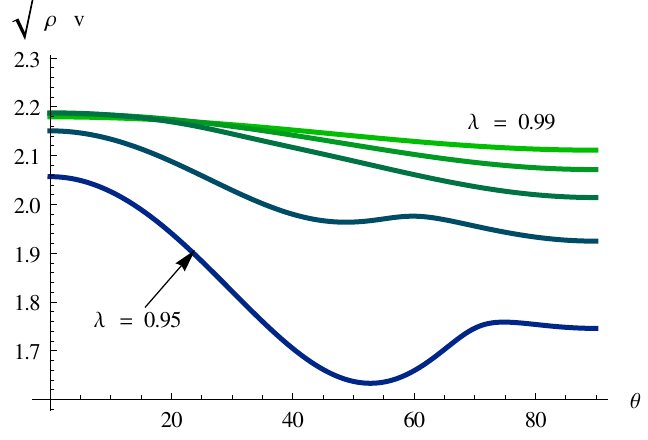}
}
\caption{Speed profiles for surface waves (plotted as $v\sqrt{\rho}$) in steel subject  to (a) uniaxial compressive stress, with pre-stretch $\lambda$ decreasing from 0.990 to 0.905 (on the left) and (b) pure shear stress, with pre-stretch $\lambda$ decreasing from 0.99 to 0.95 (on the right).
As the color of the curves  changes from green to blue, $\lambda$ is decreased by regular increments of 0.0056 from one curve to the next.}
\label{fig:steel}
\end{figure}

For deformed \emph{steel}, we observe similar characteristics for the SAW velocity profile under uniaxial compression and pure shear stress as for deformed nickel, as shown in Figure~\ref{fig:steel}. \emph{Pyrex glass} also exhibits a local minimum under pure shear stress, when $\lambda \simeq 0.975$, which then becomes a global minimum when $\lambda =0.97$, i.e. under a  compression of 3\% (figures not shown to save space).

SAWs in deformed \emph{polystyrene} behave in a more orderly way,  as they travel at their fastest along the direction of greatest stretch $ \theta= 90^\circ$ and at their slowest along $ \theta= 0^\circ$.
Although the first derivative of the velocity profile is not a monotone function of the angle, no secondary extremum develops, in contrast to the behavior of the body waves in the same material (see previous section).
\begin{figure}[!ht]
\centerline{
a)\includegraphics[height=0.22\textheight]{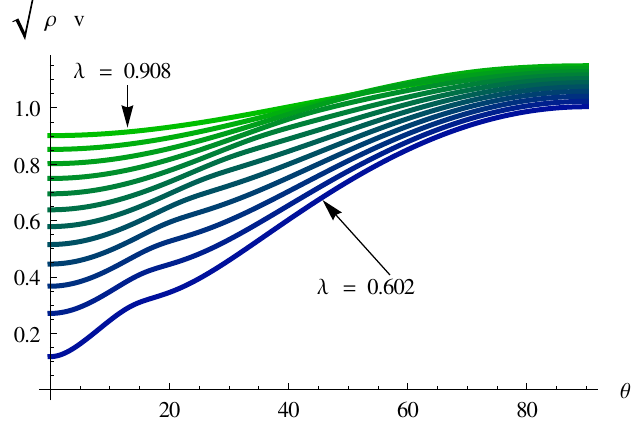}
b)\includegraphics[height=0.22\textheight]{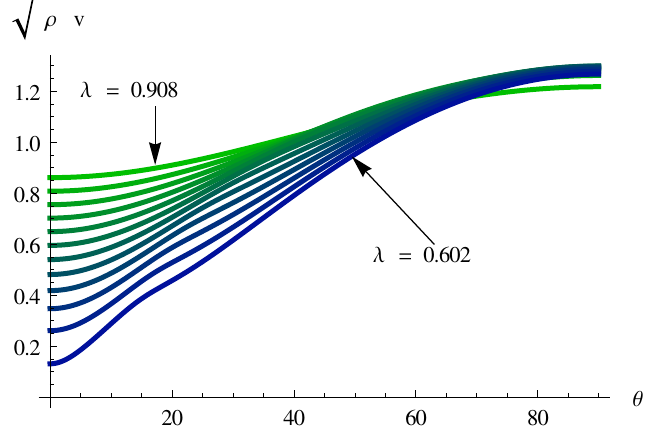}
}
\caption{Speed profiles for surface waves (plotted as $v\sqrt{\rho}$) in polystyrene subject  for (a) uniaxial compressive stress, with pre-stretch $\lambda$ decreasing from 0.908 to 0.602 (on the left) and (b) pure shear stress, with pre-stretch $\lambda$ decreasing from 0.908 to 0.602 (on the right).
As the color of the curves changes from green to blue, $\lambda$ is decreased by regular increments of 0.028 from one curve to the next.}
\label{fig:Poly}
\end{figure}

\begin{figure}[!ht]
\centerline{
a)\includegraphics[height=0.22\textheight]{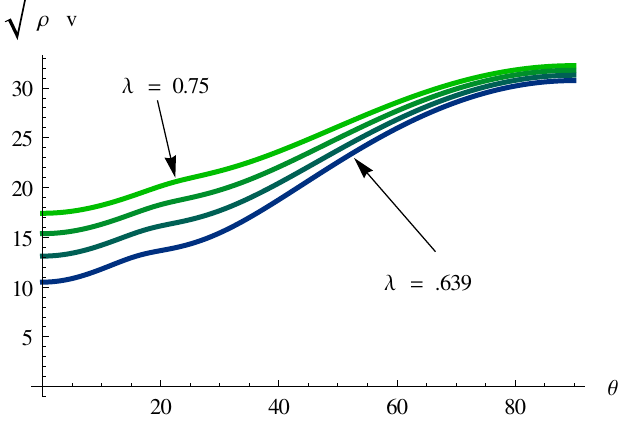}
b)\includegraphics[height=0.22\textheight]{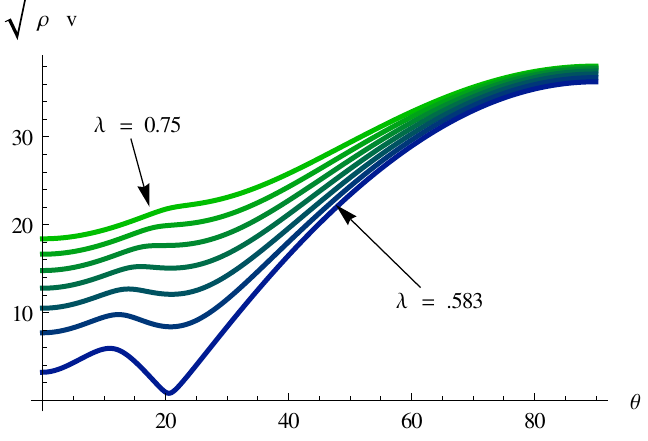}
}
\caption{Speed profiles for surface waves (plotted as $v\sqrt{\rho}$) in hydrogel subject  to (a) uniaxial compressive stress, with pre-stretch $\lambda$ decreasing from 0.750 to 0.639 (on the left) and (b) pure shear stress, with pre-stretch $\lambda$ decreasing from 0.750 to 0.583 (on the right).
As the color of the curves changes from green to blue, $\lambda$ is decreased by regular increments of 0.028 from one curve to the next.}
\label{fig:Wu}
\end{figure}

Finally, SAW propagation in deformed \emph{hydrogel} is also almost regular under uniaxial compression even at a relatively large strain (up to 40\%); see Figure~\ref{fig:Wu}(a).
However, two secondary extrema develop under pure shear stress, with the secondary minimum in an oblique direction, eventually becoming an absolute minimum; see Figure~\ref{fig:Wu}(b).


\section{Conclusion}


Clearly, the existence of oblique slowest waves greatly complicates the determination of the principal directions of strain in a deformed body.
Finding the direction where a wave travels at its slowest or fastest is not a guarantee of having determined the direction of greatest compression or tension, or that it is indeed a principal direction.
In our examples, we have found that the slowest body wave can sometimes be along an oblique direction and similarly for surface waves.
However, we found that the fastest body waves do indeed travel along a principal direction, a criterion which can thus be used to determine principal directions, at least in deformed nickel, steel, polystyrene and hydrogel.
Unfortunately, this characteristic does not carry over to the case of surface waves, as the example of nickel subject to pure shear stress shows, where the fastest surface wave is oblique.
The overall conclusion is that, for a given solid, a full analysis of wave speed variation with angle of propagation, such as that conducted in this paper, is required.


\section*{Acknowledgements}


Partial funding from a Royal Society International Joint Project grant and from the  Hardiman Scholarship programme at the National University of Ireland Galway are gratefully acknowledged.   Helpful discussions with Alexander Shuvalov are also acknowledged.



\begin{thebibliography}{99}


\bibitem[Alshits and Maugin(2005)]{AlMa05}
V.I. Alshits, G.A. Maugin. 
Dynamics of multilayers: elastic waves in an anisotropic graded or stratified plate. 
Wave Motion \textbf{41} (2005) 357--394.

\bibitem[Barnett and Lothe(1985)]{BaLo85}
D.M. Barnett, J. Lothe. Free surface (Rayleigh) waves in anisotropic elastic half-spaces: the surface impedance method,
Proc. Roy. Soc. Lond. A \textbf{402} (1985), 135--152.

\bibitem[Brillouin(1925)]{Bril25}
L. Brillouin. Sur les tensions de radiation, Ann. Phys. ser. 10 \textbf{4} (1925), 528--586.

\bibitem[Biryukov(1985)]{Biry85}
S.V. Biryukov. Impedance method in the theory of elastic surface waves, Sov. Phys. Acoust. \textbf{31} (1985), 350--354.

\bibitem[Destrade et al.(2012)]{DeMS12}
M. Destrade, J.G. Murphy, G. Saccomandi.  Simple shear is not so simple,
Int. J. Non-Linear Mech. \textbf{47} (2012), 210--214.

\bibitem[Destrade and Ogden(2013)]{DeOg13}
M. Destrade, R.W. Ogden. On stress-dependent elastic moduli and wave speeds,
IMA J. Appl. Math., in press. DOI:10.1093/imamat/hxs003

\bibitem[Destrade et al.(2005)]{DOPR05}
M. Destrade, M. Ottenio, A.V. Pichugin, G.A. Rogerson. Non-principal surface waves in deformed incompressible materials,
Int. J. Eng. Sci. \textbf{42} (2005), 1092--1106.

\bibitem[Dowaikh and Ogden(1991)]{DoOg91}
M.A. Dowaikh, R.W. Ogden. On surface waves and deformations in a compressible elastic half-space, Stability Appl. Analysis Cont. Media \textbf{1} (1991), 27--45.

\bibitem[Fu and Mielke(2002)]{FuMi02}
Y.B. Fu, A. Mielke. A new identity for the surface impedance matrix and its application to the determination of surface-wave speeds,
Proc. Roy. Soc. Lond. A \textbf{458} (2002), 2523--2543.

\bibitem[Gandhi et al.(2012)]{GaML12}
N. Gandhi, J.E. Michaels, S.J. Lee. Acoustoelastic Lamb wave propagation in biaxially stressed plates,
J. Acoust. Soc. Am. \textbf{132} (2012), 1284--1293.

\bibitem[Guyer and Johnson(2009)]{GuJo09}
R.A. Guyer, P.A. Johnson. \emph{Nonlinear Mesoscopic Elasticity}. Wiley-VCH, Weinheim (2009).

\bibitem[Hughes and Kelly(1953)]{HuKe53}
D.S. Hughes, J.L. Kelly. Second-order elastic deformation of solids, Phys. Rev. \textbf{92}  (1953), 1145--1149.

\bibitem[Kayestha et al.(2011)]{KWK11}
P. Kayestha, A.C. Wijeyewickrema, K. Kishimoto.
Wave propagation along a non-principal direction in a compressible pre-stressed elastic layer,
Int. J. Solids Struct. \textbf{48}  (2011), 2141--2153.

\bibitem[Kim and Sachse(2001)]{KiSa01}
K.Y. Kim, W. Sachse. Acoustoelasticity of elastic solids,
in \emph{Handbook of Elastic Properties of Solids, Liquids, and Gases},
Levy, Bass, Stern (Editors), \textbf{1}, 441--468. Academic Press, New York (2001).

\bibitem[Landau and Lifshitz(1986)]{LaLi86}
L.D. Landau, E.M. Lifshitz. 
\emph{Theory of Elasticity}, 3rd ed. Pergamon, New York (1986).

\bibitem[Lurie(2005)]{Lurie}
A.I. Lurie. \emph{Theory of Elasticity}. Springer, Berlin (2005).

\bibitem[Mielke and Fu(2003)]{MiFu03}
A. Mielke, Y.B. Fu. 
A proof of uniqueness of surface waves that is independent of the Stroh Formalism,
Math. Mech. Solids \textbf{9} (2003), 5--15.

\bibitem[Mihai and Goriely(2011)]{MiGo11}
L.A. Mihai, A. Goriely. 
Positive or negative Poynting effect? The role of adscititious inequalities in hyperelastic materials.
Proc. R. Soc. Lond. A \textbf{467} (2011), 3633--3646.

\bibitem[Norris(1983)]{Norr83}
A.N. Norris. 
Propagation of plane waves in a pre-stressed elastic medium,
J. Acoust. Soc. Am.  \textbf{74} (1983), 1642--1643.

\bibitem[Norris and Shuvalov(2010)]{NoSh10}
A.N. Norris, A.L. Shuvalov. 
Wave impedance matrices for cylindrically anisotropic radially inhomogeneous elastic solids.
Q. J. Mech. Appl. Math. \textbf{63} (2010), 401--435.

\bibitem[Ogden(1997)]{Ogde97}
R.W. Ogden,  \emph{Nonlinear Elastic Deformations}. Dover, New York (1997).

\bibitem[Pao et al.(1984)]{PaSa84}
Y.-H. Pao, W. Sachse,  H. Fukuoka. 
Acoustoelasticity and ultrasonic measurements of residual stresses, in W.P. Mason and R.N. Thurston, eds., \emph{Physical Acoustics}, Vol. 17, pages 61--143. Academic Press (1984).

\bibitem[Rogerson and Sandiford(1999)]{RoSa99}
G.A. Rogerson, K.J. Sandiford. 
Harmonic wave propagation along a non-principal direction in a pre-stressed elastic plate,
Int. J. Eng. Sci. \textbf{37} (1999), 1663--1691.

\bibitem[Shuvalov et al.(2004)]{ShPD04}
A.L. Shuvalov, O. Poncelet, M. Deschamps.
General formalism for plane guided waves in transversely inhomogeneous anisotropic plates,
Wave Motion \textbf{40} (2004) 413--426.

\bibitem[Tanuma et al.(2013)]{Tanu12}
K. Tanuma, C.-S. Man, W. Du. Perturbation of phase velocity of Rayleigh waves in pre-stressed anisotropic media with orthorhombic principal part,
Math. Mech. Solids, in press. DOI:10.1177/1081286512438882

\bibitem[Wu and Kirchner(2010)]{WuKi10}
M.S. Wu, H.O.K. Kirchner. Nonlinear elasticity modeling of biogels,
J. Mech. Phys. Solids \textbf{58} (2010), 300--310.

\end{thebibliography}
\end{document}